\documentclass{iopart}
\usepackage{amssymb}
\usepackage[dvips]{graphicx}

\begin{document}

\title[Searches for gravitational waves from pulsar glitches]{Searches for gravitational waves associated with pulsar glitches using a coherent network algorithm}

\author{K Hayama$^1$, S Desai$^2$, S D Mohanty$^1$, M Rakhmanov$^1$, T Summerscales$^3$, S Yoshida$^4$}
\address{$^1$The University of Texas at Brownsville, 
Brownsville, TX, 78520, USA}
\address{$^2$ Center for Gravitational Wave Physics, the Pennsylvania State University, University Park, PA 16802, USA}
\address{$^3$Department of Physics, the Andrews University, Berrien Springs, MI 49104, USA}
\address{$^4$The Southeastern Louisiana University, Hammond, LA 70402, USA}

\ead{kazuhiro.hayama@ligo.org}

\begin{abstract}
Pulsar glitches are a potential source of gravitational waves for current and future interferometric gravitational wave detectors. Some pulsar glitch events were observed by radio and X-ray telescopes during the fifth LIGO 
science run. It is expected that glitches from these same pulsars should also be seen in the future. We carried out 
Monte Carlo simulations to estimate the sensitivity of possible gravitational wave signals associated with a pulsar glitch 
using a coherent network analysis method.  We show the detection efficiency and evaluate the reconstruction accuracy of gravitational waveforms using a matched filter analysis on the estimated gravitational waveforms from the coherent analysis algorithm. 
\end{abstract}


\section{Introduction}
Pulsar glitches are a sudden spin-up of a neutron star followed by a long term relaxation period. The increase in rotational energy during such a pulsar glitch is typically $10^{43}$ ergs. 
The time scale of the sudden 
spin-up is observationally limited to within two minutes~\cite{Mcculloch:1990}. The mechanism is understood in terms of spin-down induced quakes in the neutron star's crust or the transfer of angular momentum from a superfluid in the inner crust. Some models have been proposed to explain the physical mechanism of the glitches~\cite{Baym:1969,Anderson:1975,Ruderman:1998,Takatsuka:1988,Mochizuki:1999}, however, the relevance of these models is still not certain. Though one needs more understanding of neutron star superfluidity, the drastic energy transfer should cause excitation of acoustic modes, resulting in the emission of gravitational waves. From the observations of pulsar glitches, one could infer not only the mass and radius of the stars, but also the superfluid nature, which could give us  insight into the mechanism of the pulsar glitches. According to \cite{AnderssonComer:2001}, the detection of gravitational waves caused by the superfluid modes excited by glitches gives us a good chance to understand the structure of the neutron star.

 From theoretical and observational studies, the time intervals from one glitch to the next 
from particular pulsars can be predicted~\cite{Itoh:1983}~\cite{Middleditch:2006}. The paper by Middledich et al. reports on more than seven years of monitoring PSR 0537-6910, the 16~ms pulsar in the Large Magellanic Cloud by using data acquired with the Rossi X-ray Timing Explorer. During the monitoring of this pulsar, 23 glitches were observed. From these observations, they found that the time interval from one glitch shows  a strong linear correlation with the amplitude of the first glitch, with a slope of about 6.5 days per $\mu$Hz, and  this interval can be predicted to within a few days. This predictability provides a novel opportunity for  gravitational wave detection during the future LIGO~\cite{abbottetal:2004} and VIRGO~\cite{Acerneseetal:2002} science runs. The prediction of the next glitch from an observation of a recent glitch event enables us to adjust the observation schedule for world-wide gravitational wave detectors to catch the event. This predictability 
differs from the observations of other gravitational wave sources which allow only accidental observations. Two glitches from the pulsar J0537-6910 (Aug. 4, 2006 and Nov. 25, 2005) and one glitch from the pulsar B0833-45 (Aug. 12, 2006) have been observed by radio and X-ray telescopes during the 5th LIGO science run (S5)~\cite{Middleditch:2006,Flanagan:2006}. According to figure 2 in ~\cite{Middleditch:2006}, it is quite possible that several glitch events will be observed during the next LIGO science run. 
 
We propose to look for gravitational wave signals associated with pulsar glitches using a coherent network analysis algorithm.
 Other proposed searches for gravitational waves from pulsar glitches are described in ~\cite{james:2007}. 
Unlike our coherent method, which does not assume any particular gravitational waveform, the method in~\cite{james:2007} looks for gravitational waves which have f-mode ringdown waveforms by using Bayesian model selection. 
In a coherent network analysis algorithm, one combines the data streams from multiple gravitational wave detectors coherently, taking into account the antenna patterns, geographical locations of the detectors, and the sky direction to a source. 
This method takes full advantage of the global network of interferometric gravitational wave detectors currently in operation, and improves directional 
searches, resulting in enhanced detection efficiency. Mathematically, the extraction of a gravitational wave signal is an inverse problem. Due to the degrees of freedom of a matrix of antenna patterns, the matrix becomes rank deficient at some detector networks and sky locations, resulting in the amplification of variance of the gravitational wave parameters. Another way to understand this is 
that because of the rank deficiency, one of the reconstructed polarization waveforms
becomes noisy as the noise level is dependent on the degree of the rank deficiency. To solve this ill-posed problem in the coherent network analysis, we use Tikhonov regularization~\cite{Tikhonov:1977,Rakhmanov:2006}. Besides the Tikhonov regularization, we added new time domain data conditioning~\cite{Mohanty:2002}~\cite{Mukherjee:2003} to the analysis pipeline, creating a complete stand-alone coherent network analysis pipeline called {\tt RIDGE}~\cite{Hayama:2007}.
In this paper, we demonstrate the search for gravitational waves associated with pulsar glitches using {\tt RIDGE} and show the 
detection efficiency using simulated detector noise.


\section{Overview of the {\tt RIDGE} pipeline}
{\tt RIDGE} is a  coherent network analysis method, which is described  in detail in ~\cite{Hayama:2007}. The pipeline consists of 
two main components, data conditioning and generation of detection statistics. 

The aim of the data conditioning is to whiten the data to remove frequency dependence and any instrumental artifacts. 
In {\tt RIDGE}, the data is whitened by estimating the noise floor using a running median~\cite{Mukherjee:2003}. By using a running median, the estimation of the noise floor can avoid being affected by any large outlier caused by a strong sinusoidal signal. The whitening filters are implemented using digital finite impulse response filters with the transfer function $|T(f)| = 1/\sqrt{S(f)}$, where $S(f)$ is the power spectral density (PSD) of detector noise as a function of frequency $f$. The filter coefficients are obtained by using the PSD obtained from a user-specified training data segment. The digital filter is then applied to a longer stretch of data. 

Narrow-band noise artifacts known as {\em lines} are a typical feature of the output of interferometric gravitational wave detectors. Lines originate from effects associated with the functioning of the detectors such as mirror/suspension resonant modes, power line interference during operation, calibration lines, and effects from environmental sources such as  vacuum pumps~\cite{Chassande-Mottin:2005}. These lines dominate over a specific band, which reduces the signal-to-noise ratio and the  accuracy in  recovering  gravitational wave signals. 
In {\tt RIDGE}, the whitening step above is followed by a line estimation and removal step  using a method called {\em median based line tracker} ({\tt MBLT}, for short) and is described in ~\cite{Mohanty:2002}. Essentially, the method consists of estimating the amplitude and phase modulation of a line feature at a given carrier frequency. The remarkable feature of this method is that the line removal does not affect the signals in any significant way. This is an inbuilt feature of {\tt MBLT} which uses the running median for 
the estimation of the line amplitude and phase functions. Transient signals appear as outliers in the amplitude/phase time series and are rejected by the running median estimate.

The resulting conditioned data is passed on to the next step which consists of  generation of a detection statistic. The basic 
algorithm implemented for the coherent network analysis in {\tt RIDGE} is Tikhonov regularized
 maximum likelihood, described in ~\cite{Rakhmanov:2006}. Recent studies~\cite{Klimenko+etal:2005,Rakhmanov:2006,Mohanty+etal:2006} show that the inverse problem of the response matrix of a detector network becomes an ill-posed one and the resulting variance of the solution becomes large. This comes from the rank deficiency of the detector response matrix. The strength of the rank deficiency depends on the sky location, and therefore,  plus or cross-polarized gravitational wave signals from some directions on the sky become too noisy.  In {\tt RIDGE}, we reduce this ill-posed problem by using the Tikhonov regulator, which is a function of the sky location. The input to the algorithm is a set of equal length, conditioned data segments from the detectors in a given network. The output, for a given sky location $\theta$ and $\phi$, is the value of the likelihood of the data maximized over all possible $h_+$ and $h_\times$ waveforms with durations less than or equal to the data segments.
The maximum  likelihood values are obtained as a function of $\theta$ and $\phi$ -- this two dimensional output called a 
{\em skymap}. Using the entire skymap, we construct {\it radial distance} which scales the maximum likelihood by the mean location of the same quantities in the absence of a signal. Detailed description about the detection statistic is given in ~\cite{Hayama:2007}.

\section{Simulations and results}
We performed Monte Carlo simulations to estimate the detection efficiency of possible gravitational waves from pulsar glitches. The network consisted of the 
4~km and 2~km LIGO Hanford (H1 and H2) and LIGO Livingston (L1) interferometers. For the detector noise amplitude spectral densities, we used the 
design sensitivity curves for the LIGO detectors as given in~\cite{LIGOdesignsens} and kept the locations and orientations the same as the 
real detectors. For H2, the sensitivity is $\sqrt{2}$ times less than H1. Gaussian, stationary noise was generated ($\sim 2000$~sec) by first using 3 independent realizations of white noise and then passing them through finite impulse response (FIR) filters having transfer functions that {\em approximately} match the design curves.
To simulate instrumental artifacts, we added sinusoids with large amplitudes at $(54, 60, 120, 180, 344, 349, 407, 1051)$~Hz. 

\begin{figure}
\begin{center}
\includegraphics[width=0.6\linewidth,height=0.4\linewidth]{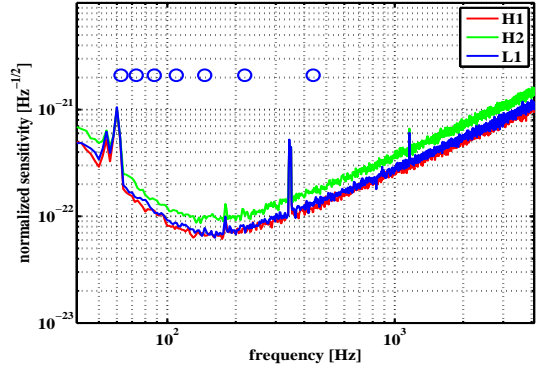}
\caption{The normalized sensitivity and root-sum-square of h of the injected ringdown signals used in the Monte Carlo simulations. The open circles are the injection signals of $h_{\rm rss}=2.1\times 10^{-21}{\rm Hz}^{-1/2}$ with center frequencies indicated in x-axis.
\label{fig:simpsds}
}
\end{center}
\end{figure}

\begin{figure}
\begin{center}
\includegraphics[width=0.6\linewidth,height=0.35\linewidth]{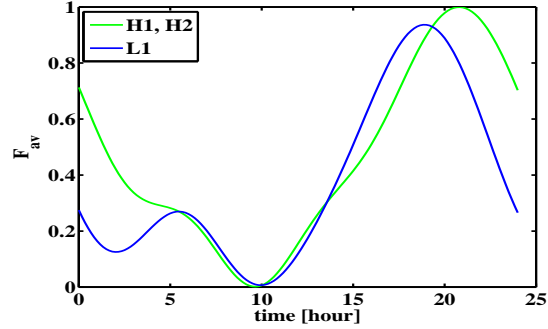}
\caption{The averaged antenna pattern, $F_{\rm av}$, at the direction to the Vela pulsar around the time when the glitch occurred at 21:59 on 12th August, 2006 which corresponds to 12 hours in x-axis in the figure. The x-axis start with preceding 12 hours from the glitch event. During the Vela pulsar glitch, the detector response is 0.157
and 0.127 for the Hanford and Livingston detectors respectively.
\label{fig:detresp}
}
\end{center}
\end{figure}

Signals of constant amplitude were added to the simulated noise at regular intervals.
The injected signals corresponded to a single source located at the right ascension 
(RA) of $8.6~{\rm hours}$ and the declination (DEC) of $-45.2~{\rm degrees}$, which was the location of Vela pulsar. We assumed that the $h_+(Q,f_c,t) = A\exp(-\pi f_ct/Q)\sin(2\pi f_ct)$
and $h_\times (Q,f_c,t) = A\exp(-\pi f_ct/Q)\cos(2\pi f_ct)$, where $t$  is time and $t\ge0$, $f_c$ is the central frequency, $Q$ the quality factor and $A$ is a constant value. The signal strength is specified in terms of root-sum-square defined as $h_{\rm rss} = \left[\int_{-\infty}^{\infty} dt\; \left(h_+^2(t) + h_\times^2(t)\right) \right]^{1/2}$. In this simulation we took $h_{\rm rss} = 2.1\times 10^{-21}{\rm Hz}^{-1/2}$, $Q=5.6$ and $f_c = 54.7,\;62.6,\;73,\;87.6,\;109.5,\;146,\;219,\;437.9$~Hz. Unfortunately, there is no reliable estimation of the amplitude yet. In this paper we use amplitudes to check the performance of this method for the ringdown with wide range of $f_c$.
Figure~\ref{fig:simpsds} shows the normalized sensitivity of the simulated detector noise for H1, H2, and  L1 along 
with the  $h_{\rm rss}$ of the simulated signals. The normalized sensitivity is defined as the amplitude spectral density divided by $(F_+^2+F_\times^2)^{1/2}$ where $F_+$, $F_\times$ are antenna patterns for $h_+$, $h_{\times}$. Though this paper uses specific ringdown waveforms with specific $f_c$, Q-value, this method can  recover any type of burst signals since this method does not assume any particular model for the signal. 
 Figure~\ref{fig:detresp} shows the averaged antenna pattern $F_{\rm av}$ defined as $F_+^2+F_\times^2$. When the Vela pulsar glitch occurred at 21:59 on 12th August, 2006, the values of the detector antenna function were 0.16 and 0.12 for the LIGO Hanford and Livingston detectors respectively.

The simulated data was generated at the sampling frequency of 16384~Hz and then passed through the data conditioning
pipeline. Besides downsampling the data to 2048~Hz by applying the same anti-aliasing filter to all data streams, 
the data conditioning pipeline applies time domain whitening 
filters that were trained, as described earlier, on the first two seconds of data for
each detector (without any injected signals). In real data analysis, we use several seconds of data, which is stationary, around 
the event excluding on-source data. On-source data refers to a data set in which the signal is likely to be present. In the case 
of the VELA pulsar glitch, which occurred on Aug. 6, 2006, the reported glitch epoch is $\simeq 100$ seconds. The delay due to 
dispersion of the radio wave signal through the interstellar medium is of the order of 100 ms and can be ignored. Therefore, 
on-source for the pulsar glitch is 100 ms of data around the event. 
Since we don't know the signal model for pulsar glitches, it is difficult to set an optimal length. So we use 0.5 seconds for one 
demonstration. Note that in real data analysis, we should use several integration lengths or clustering approach to improve 
the detection efficiency.

\subsection{ROC curve}
\begin{figure}
\begin{center}
\includegraphics[width=\linewidth]{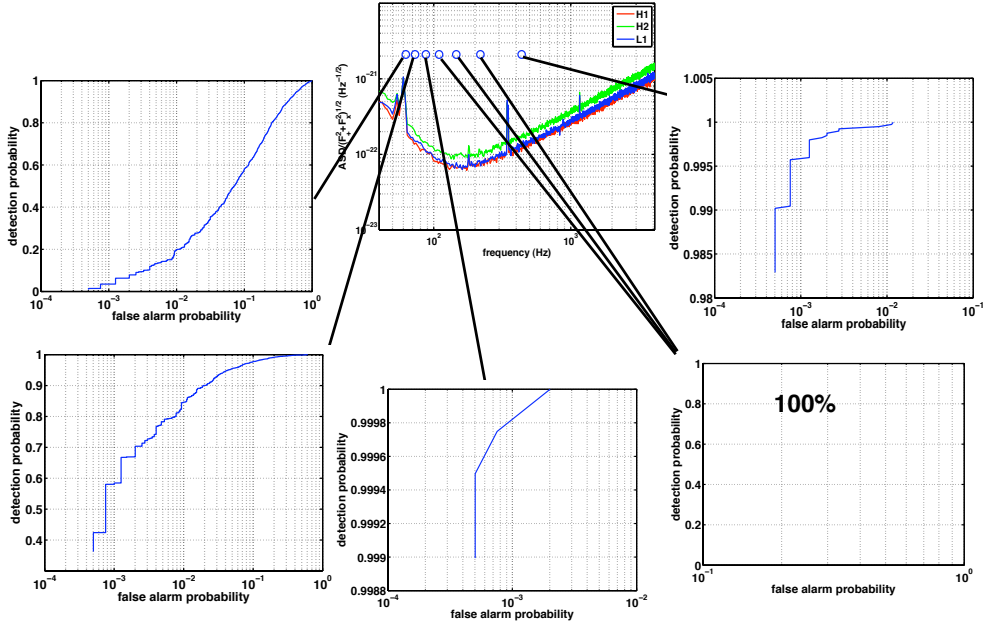}
\caption{Receiver Operating Characteristic. The sensitivity curve is the same as Figure~\ref{fig:detresp}. ${\rm h}_{\rm rss}$ of the injection ringdown signals are $2.1\times 10^{-21}{\rm Hz}^{-1/2}$ and characterized by the center frequency and Q-value $(f_c, Q)=(62.6, 5.6), (73, 5.6), (87.6, 5.6), (109.5, 5.6), (146, 5.6), (219, 5.6), (437.9, 5.6)$.  
\label{fig:roc}}
\end{center}
\end{figure}

Figure~\ref{fig:roc} shows Receiver Operating Characteristic (ROC) curves for the simulations described above. We calculated ROC curves for the 8 injection ringdown signals which are characterized by $h_{\rm rss}$, the central frequency and the Q-value shown above. The detection efficiencies at false probability $5\times10^{-3}$ are 0.99, 1, 1, 1, 1, 0.79, 0.14 from high central frequencies. In regard to the computation cost, all the processing of the simulation for one of injection signals presented here took 2 hours using a PC with 2 GHz CPU. The on-source data can be estimated $\simeq 120$ sec, and even if one hour data around the events is taken as off-source data, it's still not computationally expensive.

\subsection{Reconstructed waveform}
\begin{figure}
\begin{center}
\includegraphics[width=0.45\linewidth]{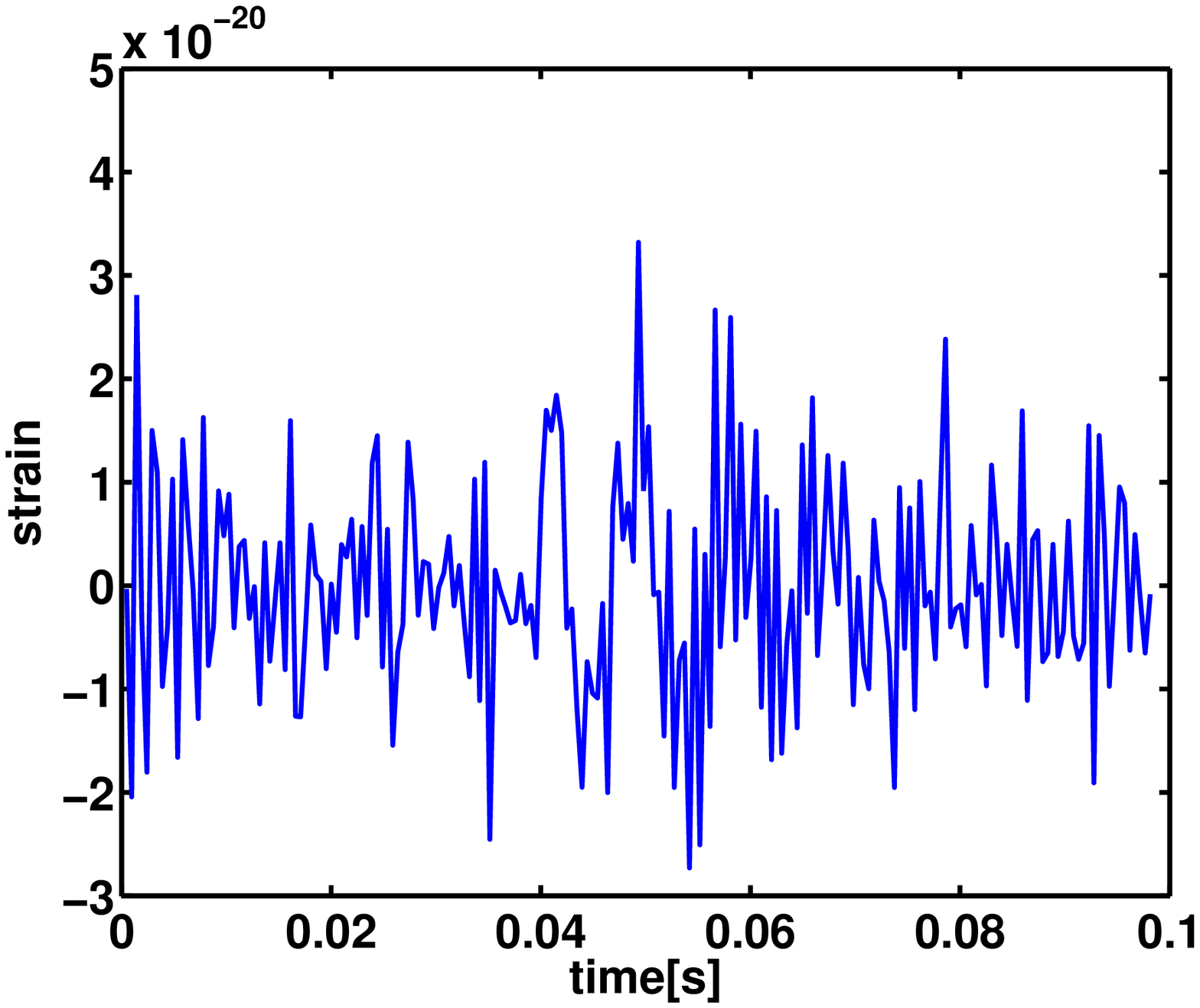}
\includegraphics[width=0.45\linewidth]{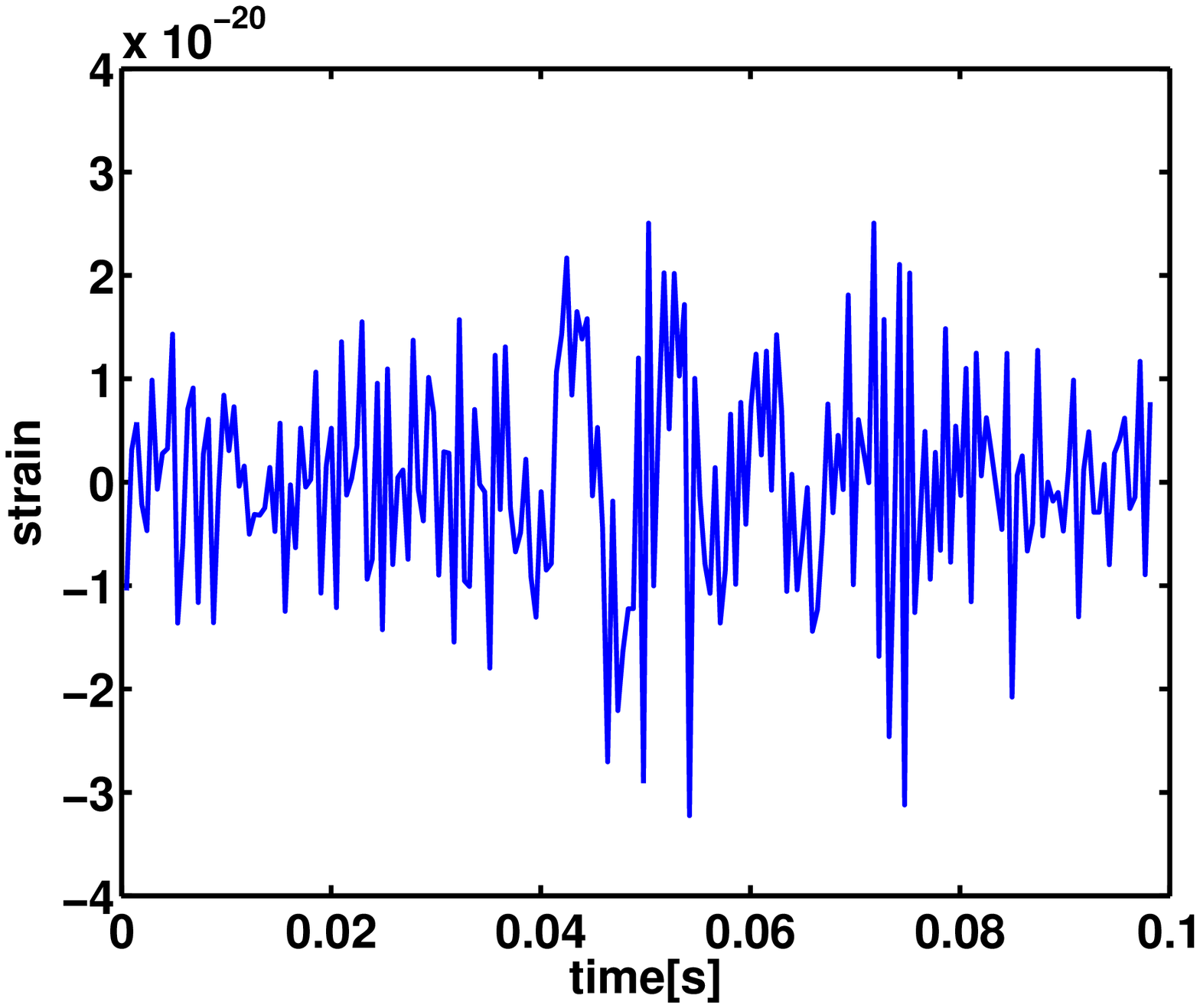}
\caption{Plots showing the reconstructed $h_+$ (left) and $h_{\times}$ (right). The corresponding injection ringdown signal is the one with a central frequency of 109.5~Hz,  Q-value of  5.6, and $h_{\rm rss}$ is $2.1\times 10^{-21}{\rm Hz}^{-1/2}$.
\label{fig:reconring}}
\end{center}
\end{figure}

Figure~\ref{fig:reconring} shows the reconstructed waveforms $h_+$, $h_\times$ for the injection ringdown signals with $f_c=109.5$~Hz. The reconstructed waveforms are subject to noise which comes from the detectors. It is necessary to de-noise the waveform when we obtain astrophysical information from unmodeled gravitational waves, and some methods for this purpose are discussed in~\cite{Hayama:2005}.

\begin{figure}
\begin{center}
\includegraphics[width=1.0\linewidth]{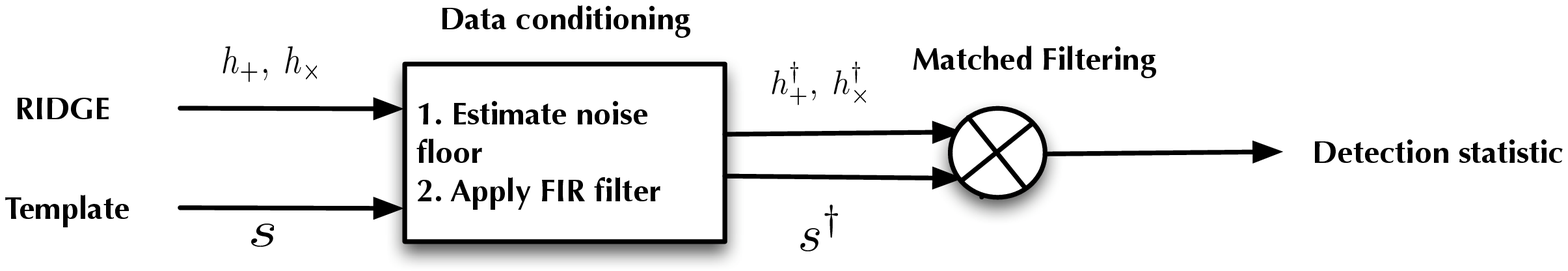}
\caption{Block diagram of the matched filter method which is applied to the reconstructed $h_+$, $h_{\times}$. The $h_+$, $h_\times$ reconstructed by the {\tt RIDGE} pipeline are passed on to the next step of whitening the reconstructed data streams. The filter coefficients are estimated using the preceding data segment, which does not contain injection signals. The template signal is also whitened by the filters with the same filter coefficients. The matched filter method is applied to the resulting whitened, reconstructed data $h_{[+,\times]}^\dagger$.
\label{fig:flowchart_matchedfilter}}
\end{center}
\end{figure}

Here, assuming the signal waveforms are known,
we carry out a matched filter method to evaluate the accuracy of the reconstruction of $h_+$, $h_\times$. Figure~\ref{fig:flowchart_matchedfilter} shows a flow chart of the pipeline for the matched filter used here. The reconstructed $h_+$, $h_\times$ are whitened by the whitening filter implemented in {\tt RIDGE}. The filter parameters are estimated using the preceding 2 seconds of data, which do not contain signals. The template $s$ is also whitened by the same whitening filter and filter parameters. The whitened data is passed on to the matched filter method which maximizes $T = {x^{\dagger}}^T s^{\dagger}/ \sigma \sqrt{{s^{\dagger}}^Ts^{\dagger}}$,
where $x^{\dagger}$ and $s^{\dagger}$ are column vectors of the whitened data $h^{\dagger}_{[+,\times]}$ and whitened template respectively. $\sigma$ is the standard deviation of the whitened data. $T$ corresponds to the signal-to-noise ratio (SNR). 

\begin{figure}
\begin{center}
\includegraphics[width=0.45\linewidth]{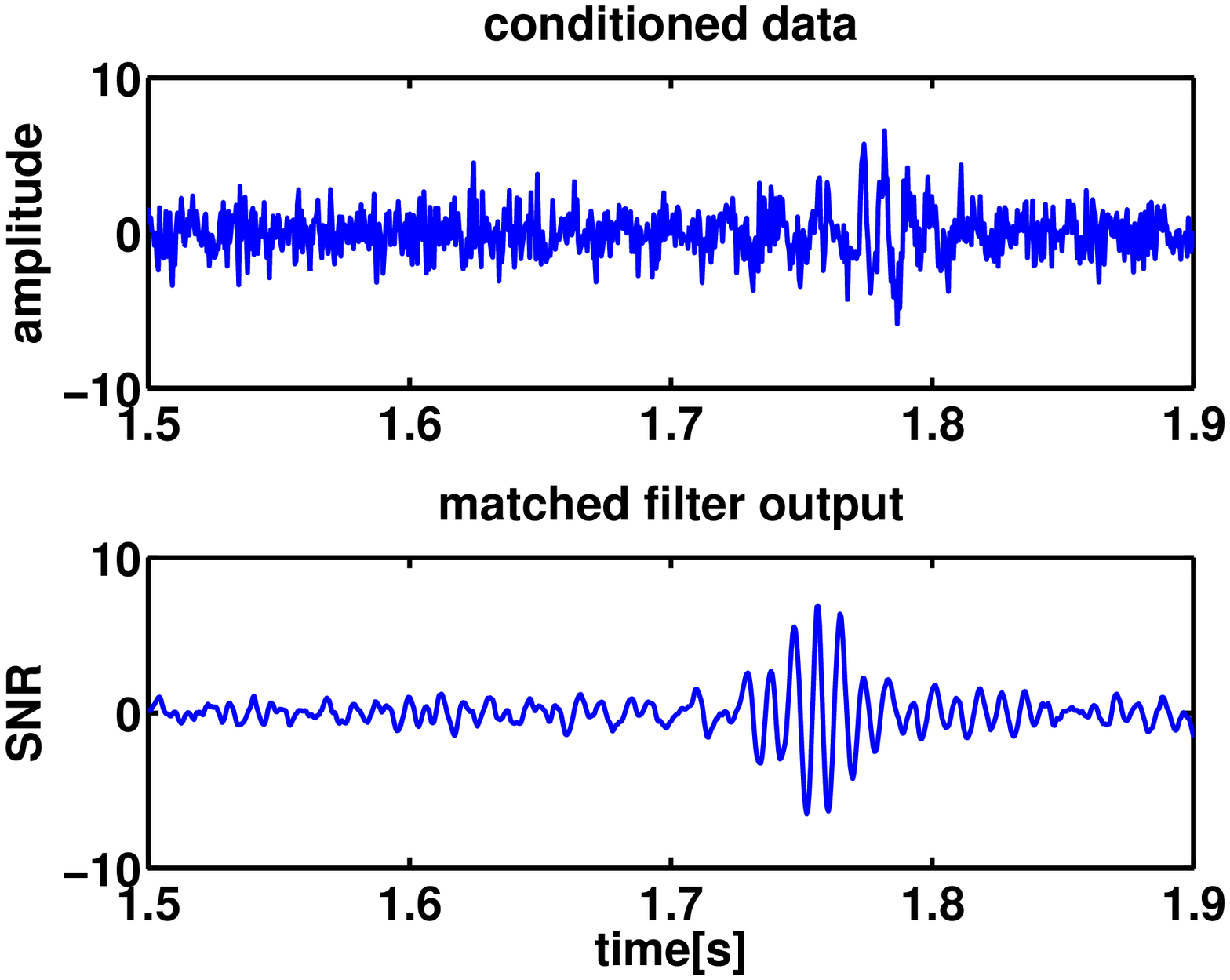}
\includegraphics[width=0.45\linewidth]{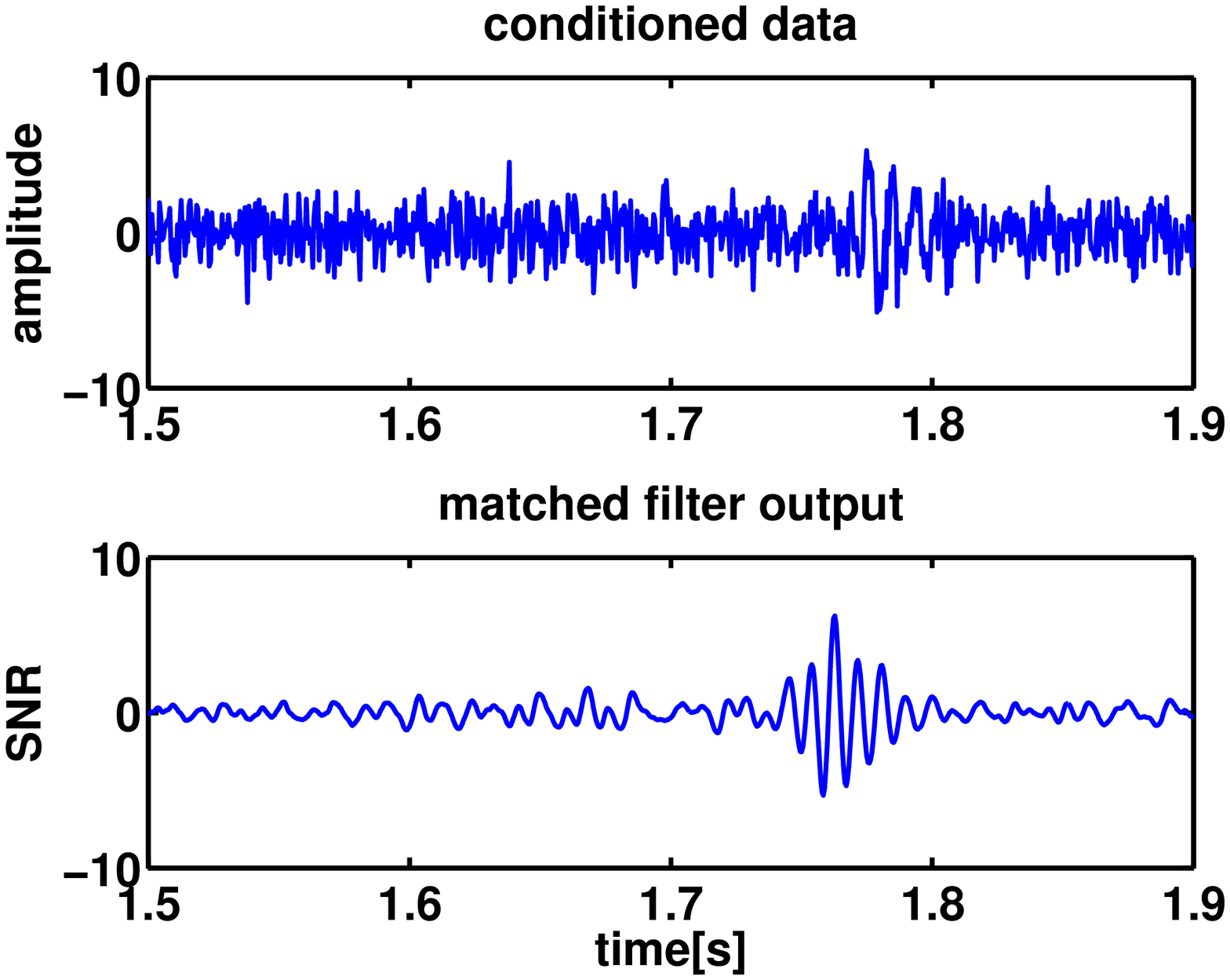} 
\caption{Upper plots show the reconstructed $h_+$ (left), $h_\times$ (right) whitened by the data conditioning routine in {\tt RIDGE}. Lower plots show the output of matched filtering for reconstructed $h_+$, $h_{\times}$ The output is the signal-to-noise ratio (SNR).
\label{fig:matchedfilter}}
\end{center}
\end{figure}

Figure~\ref{fig:matchedfilter} is the result. Upper plots show the whitened, reconstructed $h_+$, $h_\times$. Lower plots show that the SNR of the reconstructed $h_+$, $h_\times$ are 6.9, 6.3. Though any coherent network analysis has bias and we don't correct the bias, this result is not bad compared with the ideal SNRs of the injection signal: 10.1 for H1, 6.8 for H2, 8.4 for L1.


\section{Conclusion}
Pulsar glitches are a potential source of gravitational waves, and some pulsar glitches were observed during S5 by radio and X-ray telescopes. 
Our work is in progress to search for possible gravitational wave signals associated with  all the observed pulsar glitches. The next glitch from a given pulsar can be predicted from 
previous observations, which gives us a chance to adjust the observation schedule, so that the
detectors are on when the glitch occurs. This predictability of the pulsar glitches
 creates a new type of gravitational wave search as compared to other 
triggered gravitational wave searches which only allow serendipitous detections. In this paper, we perform Monte Carlo
simulations to study the detection efficiency of possible gravitational waves triggered by the Vela pulsar glitch on 12th August  2006. 
We used the  {\tt RIDGE} coherent network analysis algorithm
for this study and show the sensitivity of this method to a variety of 
possible gravitational wave signals which could happen due to this pulsar glitch. Since the 
coherent method used here can calculate the detection statistics for the 
entire sky location, it is quite easy to search for glitch events from pulsars located 
at any sky position. We also studied the accuracy of the reconstruction of gravitational 
wave signals with {\tt RIDGE} using matched filtering. The result of the matched 
filter analysis shows that the 
waveform reconstruction by {\tt RIDGE} is nearly optimal, so our 
approach can provide information about the mechanism of pulsar glitches and a clue 
to infer the inner structure of neutron stars.

\ack{We would like to thank I~Yakushin and Y~Mochizuki for fruitful discussions and valuable comments on this paper. K.H. is supported by NASA grant NAG5-13396 to the Center for Gravitational Wave Astronomy at the University of Texas at Brownsville and NSF grant NSF-HRD0734800. SDM's work was supported by NSF grant PHY-0555842. S.Y. is supported by the Southeastern Louisiana University and NSF grant PHY 0653233.
 S.D. is supported by the Center for Gravitational Wave Physics at the Pennsylvania State
 University. The Center for Gravitational Wave Physics is funded by the National Science Foundation under
 cooperative agreement PHY 01-14375. T.S. is supported by a grant from the Office of Scholarly Research at Andrews University. This paper has been assigned LIGO Document Number P080032.}
\section*{References}

\bibliographystyle{iopart-num}
\bibliography{pulsarglitch}

\end{document}